\documentclass[preprint,aps,showpacs,showkeys,onecolumn]{revtex4}%
\usepackage{amsfonts}
\usepackage{amsmath}
\usepackage{amssymb}
\usepackage{graphicx}%
\providecommand{\U}[1]{\protect \rule{.1in}{.1in}}

\begin{document}
\title{Strong coupling treatment of the polaronic system consisting of an impurity in
a condensate}
\author{W. Casteels$^{1}$, T. Van Cauteren$^{1}$, J. Tempere$^{1,2}$ and J. T.
Devreese$^{1}$}
\email{wim.casteels@ua.ac.be}
\affiliation{$^{1}$TQC, Universiteit Antwerpen, Groenenborgerlaan 171, B2020 Antwerpen, Belgium,}
\affiliation{$^{2}$Lyman Laboratory of Physics, Harvard University, Cambridge,
Massachusetts 02138, USA.}
\date[LPHYS 2010 contribution 6.5.4 (jacques.tempere@ua.ac.be)]{}

\pacs{03.75.Hh, 03.75.Hh, 71.38.Fp}
\keywords{Other Bose-Einstein condensation phenomena }
\begin{abstract}
The strong coupling treatment of the Fr\"{o}hlich-type polaronic system, based
on a canonical transformation and a standard Landau-Pekar type variational
wave function, is applied to the polaronic system consisting of an impurity in
a condensate. Within this approach the Relaxed Excited States are retrieved as
a typical polaronic feature in the energy spectrum. For these states we
calculate the corresponding effective mass and the minimal coupling constant
required for them to occur. The present approach allows to derive approximate
expressions for the transition energies between different Relaxed
Excited\ States in a much simpler way than with the full Mori-Zwanzig
approach, and with a good accuracy, which improves with increasing coupling.
The transition energies obtained here can be used as the spectroscopic
fingerprint for the experimental observation of Relaxed Excited States of
impurities in a condensate.

\end{abstract}
\maketitle

\section{Introduction}

Ultracold gasses have shown to be very useful for experimentally testing
predictions from many-body theories that arose in the context of condensed
matter physics. Their study has proven to be especially adequate for strong
coupling regimes or when there is strong correlation \cite{RevModPhys.80.885}.
Recently it has been demonstrated that if the Bogoliubov approximation is
valid the system existing of an impurity in a Bose-Einstein condensate (BEC)
can be appended to this list by a mapping to the Fr\"{o}hlich polaron system
\cite{PhysRevLett.96.210401}, \cite{PhysRevA.73.063604}. The Fr\"{o}hlich
polaron, originally introduced in condensed matter physics, is a
quasi-particle consisting of an electron with its self-induced polarization
cloud. The Fr\"{o}hlich polaron Hamiltonian has resisted exact analytical
diagonalization and has been submitted to many approximation methods (for a
detailed review of the Fr\"{o}hlich polaron in condensed mater physics we
refer to Ref. \cite{Polarons} and for more specialized topics to Ref.
\cite{BoekDevreese}). The most successful of these methods is the variational
path integral approach developed by Feynman in Ref. \cite{PhysRev.97.660}. A
polaronic coupling parameter was identified that allows to make a distinction
between three regimes: weak coupling, intermediate coupling and strong
coupling. Also the internal excitation structure of the polaron was
investigated by studying the optical absorption and in the intermediate and
strong coupling regime a resonance was found \cite{PhysRevLett.22.94},
\cite{PhysRevB.5.2367}. This resonance was associated with a transition to a
Relaxed Excited State (RES), an excitation in which the induced lattice
polarization is (self-consistently) adapted to the excited state of the
electron. More recently the polaron optical absorption was studied numerically
with diagrammatic quantum Monte Carlo numerical techniques
\cite{PhysRevLett.91.236401}. At strong coupling some differences in the
linewidth were found between the spectra derived in \cite{PhysRevB.5.2367} and
\cite{PhysRevLett.91.236401}. Recently these differences were better
understood through the introduction of the Extended Memory Function Formalism
(see Ref. \cite{PhysRevLett.96.136405}) but they have not yet been completely
clarified. Also no material is known that exhibits polaronic strong coupling.
A better understanding of the intermediate and strong coupling regimes is
needed, i.a. since polaronic effects may play an important role in
unconventional pairing mechanisms occurring in high-temperature
superconductivity \cite{SupercondBipolarons}.

In the present work we start by recalling the polaronic system consisting of
an impurity in a BEC and some standard results from the strong coupling
approximation. Subsequently, the standard strong coupling approximation is
applied to the system consisting of an impurity in a BEC. Within this
approximation a study is presented of the Relaxed Excited States and the
internal transition frequencies of the BEC-impurity polaron. These results are
compared with the more detailed study of the internal excitation spectrum of
Ref. \cite{Bragg}. Furthermore the minimal coupling parameters required for
the Relaxed Excited States to appear are deduced, and the effective masses
corresponding to these states are obtained.

\subsection{The polaronic system consisting of an impurity in a condensate}

In Ref. \cite{PhysRevB.80.184504} it is shown that if the Bogoliubov
approximation is valid the Hamiltonian of an impurity in a condensate can be
written as the sum of a mean field term and the Fr\"{o}hlich polaron
Hamiltonian $\hat{H}_{pol}$, given by:%
\begin{equation}
\widehat{H}_{pol}=\frac{\widehat{p}^{2}}{2m_{I}}+\sum_{\vec{k}}\hbar
\omega_{\vec{k}}\widehat{a}_{\vec{k}}^{\dag}\widehat{a}_{\vec{k}}+\sum
_{\vec{k}}V_{\vec{k}}e^{i\vec{k}.\widehat{\vec{r}}}\left(  \widehat{a}%
_{\vec{k}}+\widehat{a}_{-\vec{k}}^{\dag}\right)  .\label{PolHam}%
\end{equation}
In this expression the first term represents the kinetic energy of the
impurity with mass $m_{I}$ and position (momentum) operators $\widehat{r}$
($\widehat{p}$), the second term is the kinetic energy of the Bogoliubov
excitations with creation (annihilation) operator $\widehat{a}_{\vec{k}}%
^{\dag}$ ($\widehat{a}_{\vec{k}})$ and the third term describes the
interaction between the impurity and the Bogoliubov excitations. The
Bogoliubov dispersion is given by:%
\begin{equation}
\hbar \omega_{\vec{k}}=ck\sqrt{1+\left(  \xi k\right)  ^{2}/2},\label{BogDisp}%
\end{equation}
where use was made of the healing length of the condensate: $\xi=1/\sqrt{8\pi
a_{BB}n_{0}}$ and the speed of sound in the condensate: $c=\hbar/\left(
\sqrt{2}m_{B}\xi \right)  $, with $n_{0}=N_{0}/V$ the condensate density,
$a_{BB}$ the boson-boson scattering length and $m_{B}$ the mass of the bosons.
The interaction amplitude is given by:%
\begin{equation}
V_{\vec{k}}=\sqrt{N_{0}}\left[  \frac{\left(  \xi k\right)  ^{2}}{\left(  \xi
k\right)  ^{2}+2}\right]  ^{1/4}g_{IB},\label{IntAmp}%
\end{equation}
where $g_{IB}$ is the amplitude of the impurity-boson contact potential which
(at low temperature) is completely determined by the relative mass $m_{r}$ and
the impurity-boson scattering length $a_{IB}$ through $g_{IB}=2\pi \hbar
^{2}a_{IB}/m_{r}$. The Feynman variational path integral technique was applied
to this specific polaron system in Ref. \cite{PhysRevB.80.184504} and it
appeared that in this case the polaronic coupling parameter $\alpha$ can be
defined as follows:%
\begin{equation}
\alpha=\frac{a_{IB}^{2}}{a_{BB}\xi}.\label{CouplingPar}%
\end{equation}
As in the case of the solid-state Fr\"{o}hlich polaron a transition was found
between a weak coupling regime and a strong coupling regime. Furthermore it
was noted that the polaronic coupling parameter (\ref{CouplingPar}) can be
tuned externally by a magnetic field through a Feshbach resonance (see e.g.
\cite{Pitaevskii}). With this technique a coupling parameter $\alpha$ of the
order of $10$ is expected to be experimentally feasible. This suggests that by
examining an impurity in a condensate it may be possible to experimentally
reveal the internal excitation structure of the polaronic strong coupling
regime for the first time.

\subsection{Standard polaron strong coupling treatment}

In this section a summary of the polaronic strong coupling approximation is
presented which was introduced by Landau and Pekar for the description of the
ground state in Ref. \cite{LandauPekar}. These results were shown to be
accurate in the asymptotic strong coupling limit for a symmetrical, exactly
soluble one dimensional polaron model in Ref. \cite{Devreese1966196}. The
strong coupling formalism was later extended for the calculation of the energy
level of the lowest Relaxed Excited State in Refs. \cite{Pekar,Bogoliubov}.

The product Ansatz is assumed which means that the wave function of the system
$\left \vert \Psi \right \rangle $ can be written as the product of the wave
functions of the impurity $\left \vert \psi_{I}\right \rangle $ and the
Bogoliubov excitations $\left \vert \phi \right \rangle $: $\left \vert
\Psi \right \rangle =\left \vert \psi_{I}\right \rangle \left \vert \phi
\right \rangle $. The wave function $\left \vert \phi \right \rangle $ is assumed
to be of the form: $\left \vert \phi \right \rangle =\hat{U}\left \vert
0\right \rangle $, with $\left \vert 0\right \rangle $ the vacuum and $\hat{U}$ a
canonical transformation:%
\begin{equation}
\hat{U}=\exp \left[  \sum_{\vec{k}}\left(  f_{\vec{k}}\widehat{a}_{\vec{k}%
}-f_{\vec{k}}^{\ast}\widehat{a}_{\vec{k}}^{\dag}\right)  \right]  ,
\end{equation}
with $\left \{  f_{\vec{q}}\right \}  $ variational parameters. If the
expectation value of the polaron Hamiltonian with respect to $\left \vert
\Psi \right \rangle $ is minimized as a function of the variational functions
$\left \{  f_{\vec{q}}\right \}  $ the following standard expression is found
for the energy:%

\begin{equation}
E=K-\sum_{\vec{k}}\frac{\left \vert V_{\vec{k}}\right \vert ^{2}\left \vert
\rho_{\vec{k}}\right \vert ^{2}}{\hbar \omega_{\vec{k}}}.\label{VarEnerg}%
\end{equation}
where the kinetic energy $K$ and the Fourier transform of the density
$\rho_{\vec{k}}$ of the impurity were introduced:%
\begin{align}
K &  =\left \langle \psi_{I}\left \vert \frac{\widehat{p}^{2}}{2m_{I}%
}\right \vert \psi_{I}\right \rangle ;\label{KinEn}\\
\rho_{\vec{k}} &  =\left \langle \psi_{I}\left \vert e^{i\vec{k}.\widehat
{\vec{r}}}\right \vert \psi_{I}\right \rangle .\label{Dens}%
\end{align}
Expression (\ref{VarEnerg}) is strictly speaking an upper bound for the groud
state energy. When considering excitations of the impurity's wave function, it
is assumed that expression (\ref{VarEnerg}), with appropriate $\rho_{\vec{k}}%
$, gives a good approximation for the energy of the excited polaron.

\subsection{Effective mass}

The above formalism can be extended to allow for a calculation of the
strong-coupling polaron effective mass
\cite{Pekar,Bogoliubov,Vansant,Evrard1965295}. The total momentum
$\mathcal{\vec{P}}$ of the polaron system, given by $\mathcal{\vec{P}%
=}\widehat{\vec{p}}+\sum_{\vec{k}}\hbar \vec{k}\widehat{a}_{\vec{k}}^{\dag
}\widehat{a}_{\vec{k}}$, commutes with the polaron Hamiltonian (\ref{PolHam})
and is conserved. After introducing a Lagrange multiplier $\vec{v}$ the
following expression has to be minimized:%
\begin{equation}
\widehat{H}\left(  \vec{v}\right)  =\widehat{H}_{pol}-\vec{v}.\left(
\widehat{\vec{p}}+\sum_{\vec{k}}\hbar \vec{k}\widehat{a}_{\vec{k}}^{\dag
}\widehat{a}_{\vec{k}}-\mathcal{\vec{P}}\right)  ,\label{HamExt}%
\end{equation}
with $\vec{v}$ a Lagrange multiplier which physically corresponds to the
averaged velocity of the polaron. The polaron effective mass $m^{\ast}$ is
then determined from the relation between the momentum and the averaged
velocity of the polaron: $\mathcal{\vec{P}}=m^{\ast}\vec{v}$. The trial wave
function for a system in which the polaron moves with velocity $\vec{v}$ is,
i.e.:%
\begin{equation}
\left \vert \Psi \right \rangle =\exp \left[  \frac{im_{I}\vec{v}.\widehat{\vec
{r}}}{\hbar}\right]  \left \vert \psi_{I}\right \rangle \left \vert
\phi \right \rangle ,
\end{equation}
If the expectation value of $\widehat{H}\left(  \vec{v}\right)  $ is taken
with respect to $|\Psi>$ and then minimized as a function of the variational
functions $\left \{  f_{\vec{q}}\right \}  $ the following expression is found
for the energy:%
\begin{equation}
E\left(  \vec{v}\right)  =K-\frac{m_{I}v^{2}}{2}+\vec{v}.\mathcal{\vec{P}%
}-\sum_{\vec{k}}\frac{\left \vert V_{\vec{k}}\right \vert ^{2}\left \vert
\rho_{\vec{k}}\right \vert ^{2}}{\hbar \omega_{\vec{k}}-\hbar \vec{v}.\vec{k}%
}.\label{Energ}%
\end{equation}
The next step is the minimization with respect to the Lagrange multiplier
$\vec{v}$. If a Taylor expansion is performed for small $\vec{v}$ an
expression is found for the effective mass:%
\begin{equation}
m_{ij}^{\ast}=m_{I}+2\hbar^{2}\sum_{\vec{k}}\frac{\left \vert V_{\vec{k}%
}\right \vert ^{2}\left \vert \rho_{\vec{k}}\right \vert ^{2}}{\left(
\hbar \omega_{\vec{k}}\right)  ^{3}}k_{i}k_{j},\label{EffMass}%
\end{equation}
with $i,j=x,y,z$. Expression (\ref{EffMass}) results in an anisotropic
effective mass for a general density $\rho_{\vec{k}}$. Since there are no
preferred directions in the system, the physically realized effective mass is
the lowest eigenvalue of the tensor $m_{ij}^{\ast}$, since it results in the
lowest energy. 

\subsection{Variational wave function for the impurity}

In the present strong coupling approximation it is assumed that the impurity
is localized. The self-induced potential is approximated as quadratic which
means that the impurity wave function is chosen to be an eigenstate of the
harmonic oscillator:%
\begin{equation}
\psi_{n_{x},n_{y},n_{z}}\left(  \lambda,\vec{r}\right)  =\prod_{i=x,y,z}%
\left(  2^{n_{i}}n_{i}!\right)  ^{-1/2}\frac{1}{\pi^{1/4}\lambda^{1/2}%
}H_{n_{i}}\left(  \frac{i}{\lambda}\right)  \exp \left(  -\frac{r^{2}}%
{2\lambda^{2}}\right)  ,\label{HarmOscEig}%
\end{equation}
with $H_{n_{i}}$ the Hermite polynomials, $n_{i}=0,1,2,...$ and $\lambda$ the
oscillator length which is taken as a variational parameter. Physically the
oscillator length $\lambda$ can be understood as a measure of the extension of
the wave function, a small $\lambda$ corresponding to a small wave function
and thus a highly localized state. The kinetic energy (\ref{KinEn}) is given
by:%
\begin{equation}
K_{n}\left(  \lambda \right)  =\frac{1}{2}\left(  \frac{3}{2}+n\right)
\frac{\hbar^{2}}{m_{I}\lambda^{2}},
\end{equation}
with $n=n_{x}+n_{y}+n_{z}$. The Fourier transform of the density (\ref{Dens})
can also be calculated as a function of $\lambda$ and $\vec{k}$ and it turns
out to be a function of $\vec{q}=\lambda \vec{k}$, i.e. $\rho_{\vec{k}}\left(
\lambda \right)  =\rho \left(  \vec{q}\right)  $.

\section{Strong coupling approximation applied to an impurity in a condensate}

In this section we introduce the Bogoliubov dispersion (\ref{BogDisp}) and the
interaction amplitude (\ref{IntAmp}) for the polaronic system consisting of an
impurity in a condensate. If the impurity wave function is assumed to be a
harmonic oscillator eigenstate (\ref{HarmOscEig}) the following expression is
found for the energy (\ref{VarEnerg}) (with polaronic units, i.e. $\hbar
=\xi=m_{I}=1$):%
\begin{equation}
E\left(  \lambda \right)  =\frac{1}{2\lambda^{2}}\left(  \frac{3}{2}+n\right)
-\frac{\alpha \mu}{2\pi^{2}}\int d\vec{k}\frac{\rho_{\vec{k}}^{2}\left(
\lambda \right)  }{2+k^{2}},\label{VarEn}%
\end{equation}
where $\alpha$ is the polaronic coupling parameter (\ref{CouplingPar}) and
$\mu$ is given as:%
\begin{equation}
\mu=\frac{\left(  1+m_{B}\right)  ^{2}}{4m_{B}}.\label{MassInfl}%
\end{equation}
From (\ref{VarEn}), it can be seen that the effective coupling parameter is
actually a product of $\alpha$ and the mass imbalance $\mu$ between bosonic
atoms and the impurity atoms. The effective coupling constant $\alpha \mu$ can
be enhanced by choosing strongly unequal masses, making $\mu$ large. In what
follows, we study the system properties as a function of $\alpha \mu$ rather
than $\alpha$. Since the density of the harmonic oscillator eigenfunctions is
a function of $\lambda \vec{k}$ only, (\ref{VarEn}) can be rewritten as:%
\begin{equation}
E\left(  \lambda \right)  =\frac{1}{2\lambda^{2}}\left(  \frac{3}{2}+n\right)
-\frac{\alpha \mu}{2\pi^{2}\lambda}\int d\vec{q}\frac{\rho^{2}\left(  \vec
{q}\right)  }{2\lambda^{2}+q^{2}}.\label{VarEn2}%
\end{equation}
Minimization of (\ref{VarEn2}) results in the following condition on the
variational parameter $\lambda$:%
\begin{equation}
\frac{1}{\lambda^{3}}\left(  \frac{3}{2}+n\right)  =\frac{2\alpha \mu}{\pi^{2}%
}\left(  \int d\vec{q}\frac{\rho^{2}\left(  \vec{q}\right)  }{\left(
2\lambda^{2}+q^{2}\right)  ^{2}}+\frac{1}{4\lambda^{2}}\int d\vec{q}\frac
{\rho^{2}\left(  \vec{q}\right)  }{2\lambda^{2}+q^{2}}\right)  .\label{VarPar}%
\end{equation}
This equation determines a minimal value for $\alpha \mu$ below which no
solution is found, i.e. the requirement to find a solution is:%
\begin{equation}
\alpha \mu \geq \min_{\lambda \in \mathcal{R}^{+}}\left[  \frac{\pi^{2}\left(
n+3/2\right)  }{2\lambda^{3}\left(  \int d\vec{q}\frac{\rho^{2}\left(  \vec
{q}\right)  }{\left(  2\lambda^{2}+q^{2}\right)  ^{2}}+\frac{1}{4\lambda^{2}%
}\int d\vec{q}\frac{\rho^{2}\left(  \vec{q}\right)  }{2\lambda^{2}+q^{2}%
}\right)  }\right]  .\label{MinAlpha}%
\end{equation}
Values of $\alpha \mu$ not satisfying (\ref{MinAlpha}) would give
$\lambda \rightarrow \infty$ and thus a completely delocalized impurity meaning
that the self-induced trapping potential is too shallow to have this type of
wave function as a bound state. For the ground state this behavior was already
observed in Ref. \cite{PhysRevB.80.184504} where above a critical $\alpha$ a
state with a large effective mass and a localized impurity wave function was found.

Introducing the Bogoliubov dispersion (\ref{BogDisp}) and the interaction
amplitude (\ref{IntAmp}) in the effective mass (\ref{EffMass}) results in (in
polaronic units):%
\begin{equation}
m_{ij}^{\ast}=1+\frac{4\alpha \mu m_{B}^{2}\lambda}{\pi^{2}}\int d\vec{q}%
\frac{\rho^{2}\left(  \vec{q}\right)  }{\left(  2\lambda^{2}+q^{2}\right)
^{2}}\dfrac{q_{i}q_{j}}{q^{2}}.\label{EffMass2}%
\end{equation}

\section{Relaxed Excited States}

If an excited state wave function is considered for the impurity, the
resulting polaronic state is a Relaxed Excited State (RES). This corresponds
to an excitation of the impurity in the self-induced potential adapted to the
excited state's wave function. In this section the numerical solutions for the
strong-coupling approach are examined for the ground state and the three
lowest Relaxed Excited States. Polaronic units are used, i.e. $\hbar=m_{I}%
=\xi=1$.

For the ground state, $n=0$ has to be considered. The first RES corresponds to
$n=1$, i.e. one of the $n_{i}$'s is taken to be $1$. The second RES,
corresponding to $n=2$, can be realized in two distinct ways: either one
$n_{i}$ equals $2$ or two $n_{i}$'s equal $1$ which will be indicated as $2a$
and $2b$, respectively.
\begin{figure}
[ptb]
\begin{center}
\includegraphics[
height=5.5551cm,
width=10.1059cm
]%
{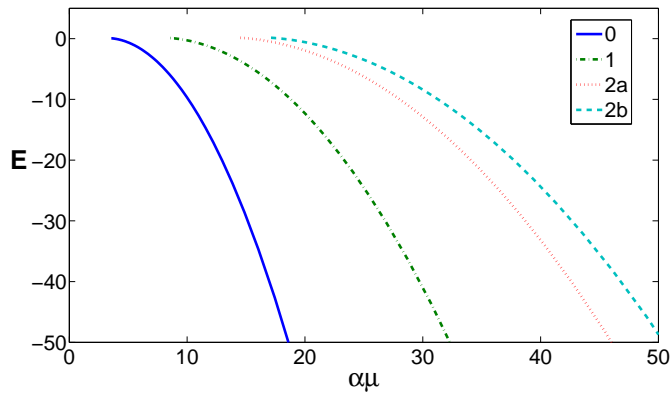}%
\caption{The variationally determined energy levels in the self-induced
potential as a function of the coupling parameter $\alpha \mu$ for the ground
state ("$0$"), for the first Relaxed Excited State ("$1$") and for the two
types of second Relaxed Excited States ("$2a$" and "$2b$").}%
\label{Fig: Energ}%
\end{center}
\end{figure}
\begin{figure}
[ptb]
\begin{center}
\includegraphics[
height=5.5531cm,
width=10.1067cm
]%
{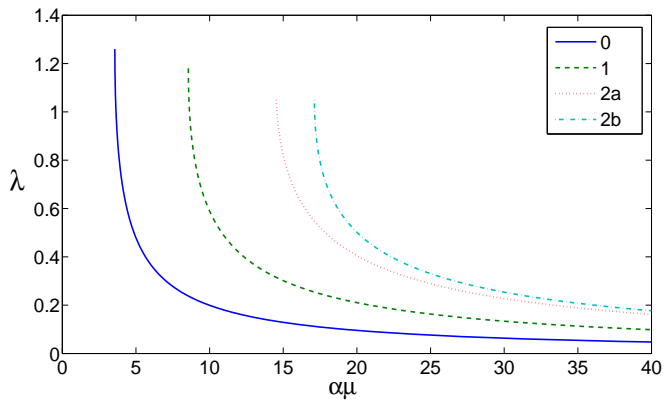}%
\caption{The variationally determined harmonic oscillator length of the
impurity wave function as a function of $\alpha \mu$ for the ground state
("$0$") and the three lowest Relaxed Excited States ("$1$", "$2a$" and
"$2b$"). }%
\label{Fig: VarPar}%
\end{center}
\end{figure}

In figure \ref{Fig: Energ} the variationally determined energy levels are
shown as a function of $\alpha \mu$ for all states discussed above and in
figure \ref{Fig: VarPar} the variational optimal oscillator length of the
impurity is shown. Note that the energies of the excited states possess no
variational upper bound character. To the left of the curves, the coupling
$\alpha \mu$ is too weak for the self-induced potential to support these
levels. For couplings just large enough for the level to exist, it is
metastable. There are, for each state, two special values of $\alpha \mu$: one
at which the self-induced potential allows for that state to exist, and one at
which the polaron energy (\ref{VarEnerg}) becomes negative, leading to a
stable state. We list these values together with the corresponding optimal
value of the variational parameter $\lambda$ in table \ref{Tab: Alfas}. From
figure \ref{Fig: VarPar} it is clear that an increase of $\alpha \mu$ results
in a decrease of $\lambda$, corresponding to a stronger self-induced potential
and thus to a more localized impurity within the polaron. Furthermore we see
that at a given $\alpha \mu$ different oscillator lengths are found for
different states which means that not only the wave function of the BEC
relaxes but also the impurity's wave function is adapted.%

\begin{table}[tbp] \centering
\begin{tabular}
[c]{l|lll|lll}
& $\left(  \alpha \mu \right)  _{exist}$ &  & $\  \  \  \lambda$ & $(\alpha
\mu)_{stable}$ &  & $\  \  \  \lambda$\\ \hline \hline
ground state & $3.56641$ &  & $1.2668$ & $3.83302$ &  & $0.8213$\\
first RES & $8.55492$ &  & $1.1850$ & $9.10277$ &  & $0.7646$\\
second RES (2a) & $14.4869$ &  & $1.1495$ & $15.3764$ &  & $0.7400$\\
second RES (2b) & $17.1152$ &  & $1.0570$ & $18.0197$ &  & $0.6852$%
\end{tabular}
\caption{The values of $\alpha \mu$ above which the self-induced potential supports the indicated levels is
given, along with the corresponding values of the variational parameter $\lambda$ (cf. eq. (19)). The last
two columns give the values of $\alpha \mu$ (and the corresponding $\lambda$) above which the indicated
levels become stable.}\label{Tab: Alfas}%
\end{table}%

\begin{figure}
[ptb]
\begin{center}
\includegraphics[
height=5.0998cm,
width=10.1059cm
]%
{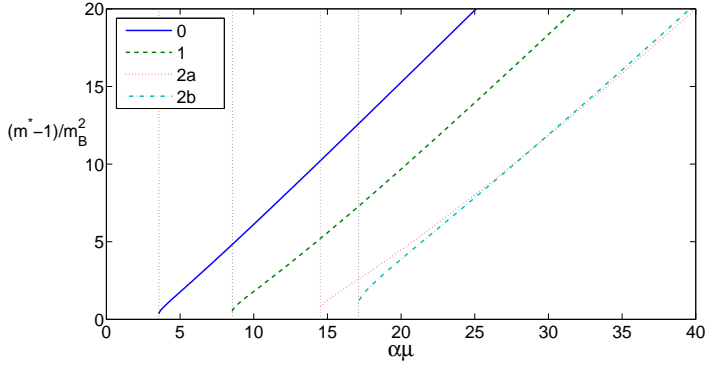}%
\caption{This figure shows $(m^{\ast}-1)/m_{B}^{2}$ where the effective mass
$m^{\ast}$ is given by eq. (20) for the ground state ("0") and for the three
lowest Relaxed Excited States ("$1$", "$2a$" and "$2b$"). The values of
$\left(  \alpha \mu \right)  $ at which the self-induced potential is deep
enough to support these states are also indicated (cf. Table \ref{Tab: Alfas})
by dotted vertical lines.}%
\label{Fig: EffMAss}%
\end{center}
\end{figure}

In figure \ref{Fig: EffMAss}, $(m^{\ast}-1)/m_{B}^{2}$ is shown, where
$m^{\ast}$ is the effective mass for the states under consideration. The
largest effective mass is found for the ground state and it is observed that
the effective masses corresponding to the two possibilities for the second RES
cross at a coupling parameter $\alpha \mu \approx28.7$.%
\begin{figure}
[ptb]
\begin{center}
\includegraphics[
height=5.5551cm,
width=10.1059cm
]%
{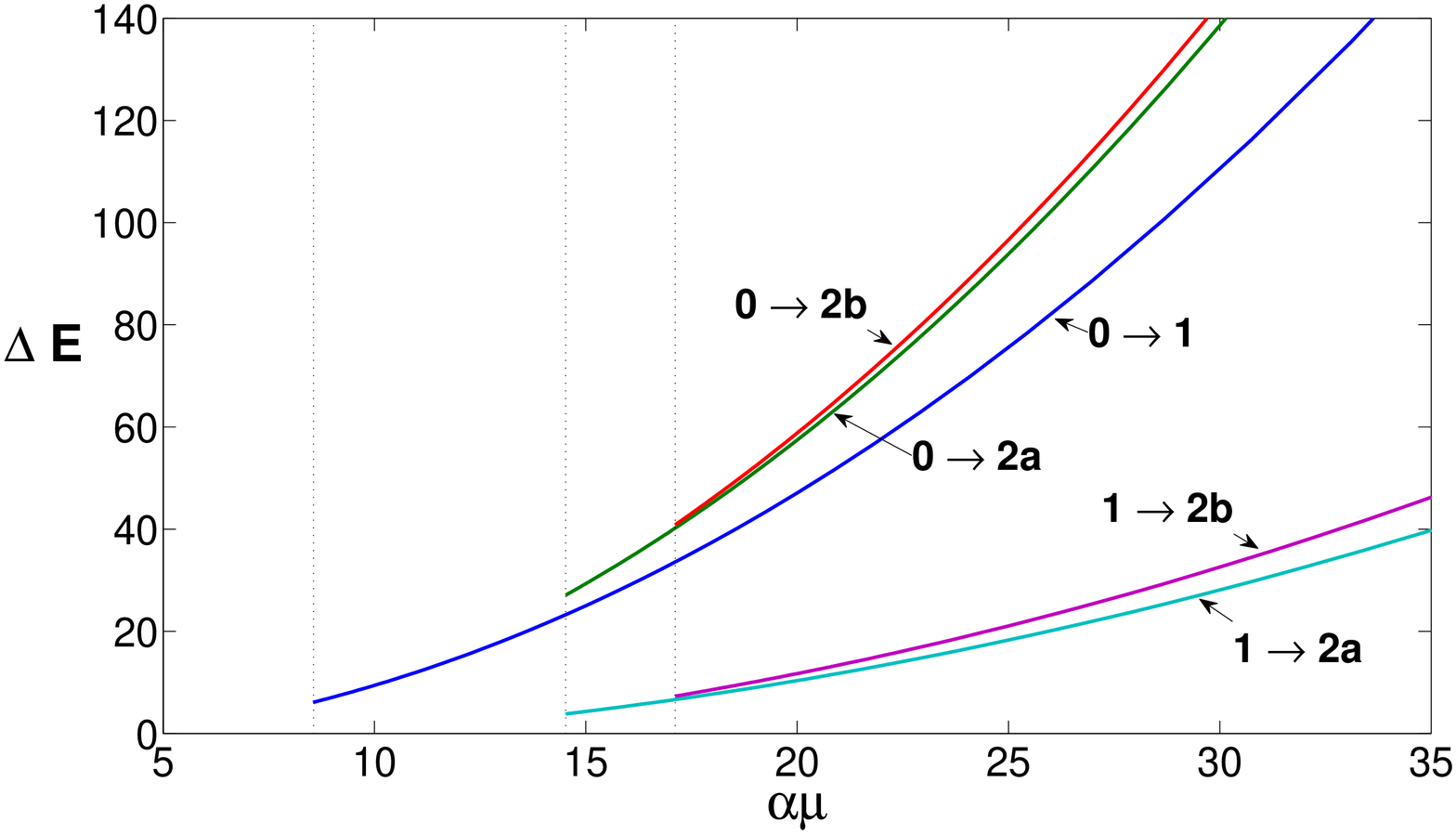}%
\caption{Transition energies from the ground state ("$0$") to the Relaxed
Excited States ("$1$", "$2a$" and "$2b$") and from the first RES ("$1$") to
the two types of the second RES ("$2a$" and "$2b$"). The critical coupling
parameters $\left(  \alpha \mu \right)  _{exist}$ from table \ref{Tab: Alfas}
for the Relaxed Excited States are also indicated by dotted vertical lines.}%
\label{Fig: TransFreq}%
\end{center}
\end{figure}
In figure \ref{Fig: TransFreq} the transition energies are shown for the
transition from the ground state to the Relaxed Excites States under
consideration and from the first RES to the two types of the second RES. For
larger $\alpha \mu$ more Relaxed Excited States are expected to exist and as a
result there will be more possible transition energies.

Recently a more detailed study of the excitation structure of the polaron
system consisting of an impurity in a condensate was performed by calculating
the response of the system to Bragg spectroscopy within the Mori-Zwanzig
projection operator formalism \cite{Bragg}. In this case the system consisting
of a lithium impurity in a sodium condensate was examined which results for
the present treatment in $\mu=1.5209$ for the mass parameter (\ref{MassInfl}).
With this value of $\mu$ the present theory is able to describe the strong
coupling ground state from $\alpha \gtrapprox2.5$. The self-induced potential
becomes deep enough to support a first Relaxed Excited State at $\alpha
\approx6$. In Ref. \cite{Bragg} the transition to the first Relaxed Excited
State is already seen at $\alpha=4$ and for larger $\alpha$ a larger
transition frequency is found, a behavior which is also observed in figure
\ref{Fig: TransFreq}. This shows that the lowest $\alpha$ at which the Relaxed
Excited State appears is slightly smaller in the treatment of Ref.
\cite{Bragg} and somewhat larger in the present strong coupling treatment. A
comparison of the transition frequency from the ground state to the first
Relaxed Excited State can be made for $\alpha$ larger than $6$. At $\alpha=8$
this results in a difference of about $10\%$ and at $\alpha=10$ a deviation of
about $5\%$ is found. This trend of a better correspondence between the
present strong coupling approximation and the arbitrary coupling treatment of
Ref. \cite{Bragg} at higher coupling parameter is consistent since the present
strong coupling theory is expected to be more accurate at larger $\alpha$.

In the present strong coupling calculation the effect of the mass imbalance is
contained solely in the factor $\mu$ which grows as the masses differ more,
leading to a larger $\alpha \mu$. This observation provides an additional tool
for facilitating the probing of the strong coupling regime, namely choosing
the masses as different as possible: a lithium impurity in a rubidium
condensate will correspond to a more strongly interacting system at the same
(Feshbach adapted) scattering lengths than a lithium impurity in a sodium condensate.

\section{Conclusions}

In this work the standard strong coupling approach to polaron theory is
applied to investigate the strong coupling regime of the polaronic system
consisting of an impurity in a Bose-Einstein condensate. Within this formalism
the critical coupling parameters required for Relaxed Excited States to appear
were deduced, and the effective masses of these states were calculated. For
impurity atoms in a condensate, these states can be experimentally most easily
probed by laser spectroscopy. The calculation of the transition energies
between different Relaxed Excited States presented here offers a
straightforward way to estimate the excitation spectrum. Comparison with the
path-integral results for the spectral density obtained within the
Mori-Zwanzig framework (see Ref. \cite{Bragg}) shows that the current method
quickly becomes very accurate (%
$<$%
5\% for $\alpha \gtrsim10$) as the coupling strength grows. Furthermore it was
shown that the polaronic strong coupling regime is easier to reach when the
masses of the bosons and the impurity differ more from each other.

\end{document}